# Logic and Reduction Operation based Hardware Trojans in Digital Design

1st Mayukhmali Das
mayukhmalidas322@gmail.com

2nd Sounak Dutta
sounak04@gmail.com

3rd Prof. Sayan Chatterjee
sayan.chatterjee@jadavpuruniversity.in

Department of Electronics and Telecommunication Engineering, Jadavpur University, India

**Abstract:** *In this paper, we will demonstrate Hardware Trojan Attacks on four different digital designs implemented on FPGA. The hardware trojan is activated based on special logical and reduction-based operations on vectors which makes the trojan-activity as silent and effective as possible. In this paper, we have introduced 5 novel trojan attack methodologies.*

## 1. Introduction

With the growing invasiveness of day-to-day gadgets, the risk of attacks on the common hardware chips and SOCs has become very much rampant. Various Hardware Trojan categories are discussed in [1]. Hardware Trojan mitigation measures are provided in [2]. [3] demonstrates how hardware trojans can be introduced in Asynchronous FIFO Buffers. Literature [4] provides interesting insights into how hardware trojan and fault attacks can lead to the degradation of accuracy in FPGA-based neural networks.

In this paper, all of the trojan attacks are performed on designs deployed in FPGA Xilinx XC7A100T-1CSG324C. This paper considers that the attacker has access to the RTL Code for each of the designs and no black box attack is being performed. We have used Quartus Prime Version 20.1.1 and ModelSim 2020.02 for Simulation and Verification Purposes.

## 2. Digital Designs

### A. 8-bit Dual Edge Detector

One of the main vulnerabilities that can be exploited in a digital design using a Flip Flop is setting up a metastable state. Since Static Timing Analysis is done with respect to edges of the clock pulse, so detection of the clock edge is very important to the inherent functioning of a digital circuit. A Dual Edge Detector detects both the positive and negative edges of a digital signal. We will now discuss how to introduce a trojan in an 8-bit dual edge detector implemented using an FPGA. The basic code for an 8-bit dual-edge detector is shown below:

```
module edge_detector ( input clock, input [7:0] in, output [7:0] dual_edge );
   reg [7:0] q;
   reg trojan;
   assign trojan = ^dual_edge;
   always@(posedge clock)
     begin
       dual_edge <= q^in ^trojan;
       q <= in;
     end
endmodule
```

The bold part of the code is how the trojan is introduced. In the beginning, the design works absolutely fine, but as soon as the **reduction operation** on the dual-edge detector register gives logic one; it complements the entire dual-edge detector vector, causing malfunction in the design and we get an error in our output. The number of samples considered for simulation was 227. Table I shows that the XOR reduction operation has a good balance between the invisibility of the trojan and fault injection. This trojan can be made more non-invasive and invisible by letting it trigger after the N-th time the reduction operation gives logic one. This can be designed using a counter. Such a trojan will be a hybrid of a Sequential (Time Bomb) Trojan [1] and a Combinational Trojan.

| Reduction operator | Number of errors in output |
|---|---|
| & | 0   (0%) |
| \| | 212 (93.4%) |
| ^ | 104 (45%) |
| ~& | 224 (98.6%) |
| ~\| | 14   (6.2%) |
| ~^ | 92   (40.5%) |

Table I. Number of errors with respect to various reduction operation

### B. 32-bit LFSR

We use Linear Feedback Shift Register in important domains like cryptography and communication. We will design a **logic operation** based trojan by exploiting the basic vulnerability of LFSR which is the all-zero state. The basic algorithm behind this hardware trojan is that we will introduce an all-zero state whenever the LFSR is being reset and at that time one particular bit of the shift register is logic one.

| Index of the bit using which the trojan is introduced | Number of errors |
|---|---|
| 5 | 2 |
| 10 | 20 |
| 18 | 18 |
| 28 | 400028 |

Table II. Variation in Error with respect to bit index at which all zero state is introduced

For simulation, we have taken 32-bit Galois LFSR with taps at 1, 2, 22, and 32. The sample size taken for simulation is 400411. We can see from Table II that the number of errors varies as we use different indexes to introduce the trojan. A higher number of mismatches indicate that we are introducing the trojan early. An exploiter should keep in mind to use that particular bit of the LFSR, which injects the trojan after a copious amount of time. Here we will prefer to introduce the trojan at bit 10 over bit 28 as the former injects the all-zero state after a long time.

### C. Mouse Controller

PS/2 Mouse is popular due to their low latency and ability to register multiple triggers at the same time, that is it can identify when both the left and the right mouse buttons are pressed simultaneously. It is evident from Table III, that PS/2 mouse data packets are 3 bytes long. Now, the PS/2 mouse protocol is relatively simple and has only a start bit check. The start bit, in this case, is the third bit of the first byte. We are proposing two different hardware trojans. The first type will inject error into the mouse data packets, and the second one will attack the design by introducing a malicious state.

| | Bit7 | Bit6 | Bit5 | Bit4 | Bit3 | Bit2 | Bit1 | Bit0 |
|---|---|---|---|---|---|---|---|---|
| Byte 0 | Y overflow | X overflow | Y sign | X sign | Always 1 | Middle Button | Right Button | Left Button |
| Byte 1 | X movement ||||||||
| Byte 2 | Y movement ||||||||

Table III.  PS/2 mouse data packets

| Reduction operation | Trojan 1 No. of Errors | Trojan 2 No. of Errors |
|---|---|---|
| & | 0 (0%) | 0 (0%) |
| \| | 196 (13.8%) | 432 (30.4%) |
| ^ | 102 (7%) | 430 (30.3%) |
| ~& | 196 (13.8%) | 432 (30.4%) |
| ~\| | 0 (0%) | 0 (0%) |
| ~^ | 94 (6.63%) | 422 (29.7%) |

Table IV. Trojan 1 and Trojan 2 Error Rates w.r.t various Reduction Operation

| Duplication at Bit number | Number of errors in the value of Received Data Packet |
|---|---|
| 1 | 2 |
| 2 | 0 |
| 3 | 0 |
| 4 | 2 |
| 5 | 2 |
| 6 | 2 |

Table V. Variation of the number of errors in the received data with respect to data bit which is being duplicated

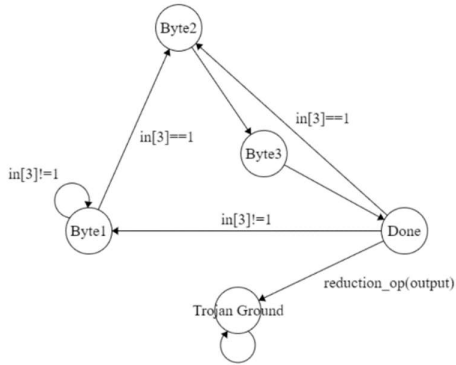

Fig. 1. State Diagram for PS/2 Mouse Controller

1. **Trojan 1:** This Trojan swaps the values of the Right and Left mouse buttons at a particular instance. We have decided that this swapping of Bit1 (Right) and Bit 0 (Left) will be governed by *reduction operation* on the mouse data packet. Variations can be done by swapping the X movement, Y movement, etc.

2. **Trojan 2:** In Figure 1, we can see the state machine diagram for the mouse controller. In[3] denotes the third bit of the current input byte. Only In[3] of the first of the three bytes is significant as it is the start bit. Now, we are introducing a new malicious state called *Trojan Ground*. The state machine will be stuck in an infinite loop as soon as the *reduction operation* on the mouse data packet gives logic 1, provided it was previously in the "Done" state as per Fig. 1. This trojan will make the mouse not function anymore until it is reset.

The number of samples taken for simulation is 1419. When we use Trojan 1, there is only error in the value of the received packet as it only modifies the data packet. On the other hand, when we use Trojan 2, it makes the mouse stop working, so the errors given in Table IV denote both start bit validation errors as well as errors in the value of the data packet. Also, from Table IV, we can clearly see that Trojan 1 is much more invisible and less invasive when compared to Trojan 2.

## D. Serial Receiver for RS232 and UART

The most common data transmission format for Serial Communication is a start bit, 8 data bits, parity bit, and stop bit. We will introduce a special kind of trojan that will duplicate[4] a particular data-bit of the received 8 data bits. We will resort to a simple approach of introducing silent data-bit duplication using a malicious state. Also, instead of data duplication for a single time, we can copy a particular data bit N-times. Figure 2 shows the state diagram of the entire design with a malicious trojan state. Using this state machine, we can find out a) whether the data packet is valid or not and b) the value of the data packet. The value of the 8-bit data packet is maintained in an output shift register.

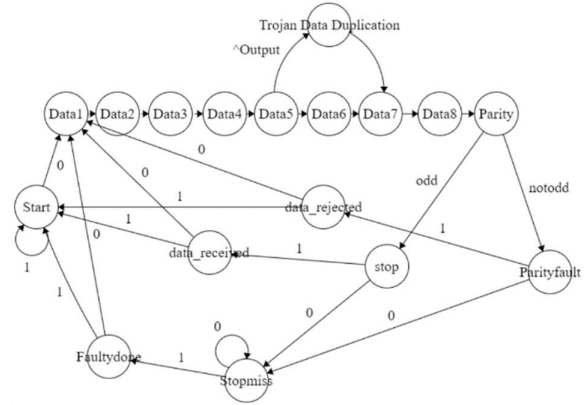

Fig. 2. State Diagram for Serial Receiver

New values are shifted in as the state machine goes through the states from *Data1* to *Data 8*. Now, the trojan will duplicate, say, the 5$^{th}$ bit of the 8 data bits, only when the *reduction operation* XOR of the current value in the output shift register is 1; or say it may be triggered if there was a simultaneous parity fault and stop bit miss in the previous data packet and so on. There can be several triggering factors for the trojan. Now when the trojan is activated and the state machine goes into the *Trojan Data Duplication* state, the shift register shifts in the previous value, instead of the new input bit, hence resulting in data duplication. The number of samples taken for simulation is 1315. We have kept the condition of activation of the trojan to be whenever the *reduction operation* XOR on the output shift register gives 1. Table V shows the error rates; that is the error in the received data. There will also be errors in the validation of the data, as parity is getting changed whenever data bits are changed. It is evident that we are getting fairly similar error rates irrespective of which bit of the packet is duplicated.

ACKNOWLEDGMENT

This work was supported by SMDP Lab, Jadavpur University. Testbenches used in this paper were generated from HDLBits website.

REFERENCES

[1] Chakraborty, Rajat Subhra, Seetharam Narasimhan, and Swarup Bhunia. "Hardware Trojan: Threats and emerging solutions." 2009 IEEE International high level design validation and test workshop. IEEE, 2009.

[2] Bhunia, Swarup, et al. "Hardware Trojan attacks: Threat analysis and countermeasures." Proceedings of the IEEE 102.8 (2014): 1229-1247.

[3] Hasan, Syed Rafay, et al. "Hardware trojans in asynchronous fifo-buffers: From clock domain crossing perspective." 2015 IEEE 58th International Midwest Symposium on Circuits and Systems (MWSCAS). IEEE, 2015.

[4] Rakin, Adnan Siraj, et al. "{Deep-Dup}: An adversarial weight duplication attack framework to crush deep neural network in {Multi-Tenant}{FPGA}." 30th USENIX Security Symposium (USENIX Security 21). 2021